\documentclass[twocolumn,showpacs,amsmath,amssymb,aps]{revtex4}
\usepackage{graphicx}
\begin{document}

\title{Solvable models for the gamma deformation having X(5)as limiting  symmetry. Removing some drawbacks of the existent descriptions}

\author{A. C. Gheorghe $^{a)}$, A. A. Raduta $^{a), b), c)}$ and Amand Faessler $^{c)}$ }

\address{$^{a)}$Department of Theoretical Physics,
Institute of Physics and Nuclear Engineering, Bucharest, POBox MG6,
 Romania}

\address{$^{b)}$Department of Theoretical Physics and Mathematics,
Bucharest University, POBox MG11, Romania}

\address{$^{c)}$Institut fuer Theoretische Physik der Universitaet Tuebingen,
Auf der Morgenstelle 14, Germany}

\date{\today}

\begin{abstract}
Two solvable Hamiltonians for describing the dynamic gamma deformation,
are proposed. The limiting case of each of them is the $X(5)$ Hamiltonian.
Analytical solutions for both energies and wave functions, which are periodic in $\gamma$, are presented in terms
of spheroidal and Mathieu functions, respectively. Moreover, the gamma
depending factors of the transition operator can be treated.
\end{abstract}

\maketitle

Phenomenological formalisms like liquid drop (LD) \cite{Bo}, rotation-vibration
\cite{GrFa}, gamma unstable (GU) \cite{Jean}, triaxial rotor (TR) \cite{Filip} Greiner-Gneuss (GG) \cite{Gneus}, interacting boson (IBA) \cite{Iache,Iache1}, coherent state model \cite{Rad1}, interpret some of the existent data by referring to a nuclear equilibrium shape, defining a nuclear phase with specific properties.
It has been noticed that a given nuclear phase may be
associated to a certain symmetry.
Thus, the gamma unstable nuclei can be described by the $O(6)$ symmetry
\cite{Jean}, the gamma triaxial nuclei by the rigid triaxial rotor $D2$ symmetry
\cite{Filip}, the symmetric rotor by the $SU(3)$ symmetry and the spherical vibrator by the $U(5)$ symmetry.

IBA succeeded to describe the basic properties of low lying states in a large
number of nuclei in terms of the symmetries associated to the system of
quadrupole ($d$) and
monopole ($s$) bosons  which  generate a $U(6)$ algebra. The three
limiting symmetries $U(5)$, $O(6)$, $SU(3)$ mentioned above, are dynamic symmetries for $U(6)$. Moreover, for each of these symmetries a specific group reduction chain  provides the quantum numbers characterising the states, which are suitable for a certain region of nuclei.
A nice classification scheme was provided by Casten \cite{Cast}, who placed all
nuclei on the border of a symmetry triangle. The vertices of this triangle symbolise the $U(5)$ (vibrator), $O(6)$ (gamma soft) and $SU(3)$ (symmetric rotor), while the legs of the triangle denote the transitional region. In Ref. \cite{Gino,Diep}, it has been proved
that on the $U(5)-O(6)$ transition leg there exists a critical point
for a second order phase transition while the $U(5)-SU(3)$ leg has a first order phase transition. 

Recently, Iachello \cite{Iache2,Iache9} pointed out that these critical points
correspond to distinct symmetries, namely $E(5)$ and $X(5)$, respectively. 
Remarkable is the fact that the experimentalists  found representatives for
the two
symmetries \cite{Zam,Zam1}.
Short after the pioneering papers concerning critical point  symmetries
appeared, some other attempts have been performed, using other potentials like Coulomb, Kratzer \cite{Fort} and
Davidson potentials \cite{Bona}. These potentials yield also Schr\"{o}dinger solvable equations and the corresponding results may be interpreted in terms of
symmetry groups.

The departure from the gamma unstable picture has been treated by several
authors whose contributions are reviewed by Fortunato in Ref. \cite{Fortu2}.
The difficulty in treating the gamma degree of freedom consists in the fact
that this variable is coupled to the rotation variables.
A full solution for the Bohr-Mottelson Hamiltonian including an explicit
treatment of gamma deformation variable can be found in Ref. \cite{Ghe1,Rad2}.
Therein we treated separately also the gamma unstable and rotator Hamiltonians. A more complete study of the rotor Hamiltonian and the distinct phases associate to a tilted moving rotator is given in Ref.
\cite{Ghe2}. In the recent publications, for the sake of simplicity, one uses
model potentials which are sums of a beta and a gamma depending potentials.
In this way the nice feature for the beta variable to be decoupled from the
remaining 4 variables, specific to the harmonic liquid drop, is preserved.
Further, the potential in gamma is expanded either around to $\gamma=0$ or
around $\gamma=\pi/6$.
In the first case, if only the singular term is retained one obtains the infinite square well model described by Bessel functions in gamma. If the $\gamma^2$ term is added to this term, the Laguerre functions are the eigenstates of the approximated gamma depending Hamiltonian, which results in defining the representation of the $X(5)$ symmetry group. The drawback of these approximation consists in that the resulting function are not periodic as required by the starting 
Hamiltonian. Moreover, they are orthonormalized on unbound intervals although the underlying equation was derived under the condition of $|\gamma|$ small. The scalar product for  the space of the resulting functions is not defined with the measure $|\sin3\gamma| d\gamma$ as  happens in the LD model. Under these circumstances the approximated Hamiltonian in 
$\gamma$ is not Hermitian.

In the present letter we offer two solutions which cure the mentioned drawbacks. One is based on spheroidal functions while the other one on the Mathieu equation.
These cases will be treated separately. However, in order to stress on the importance of curing the  bad behaviour of the gamma depending wave function we start by saying few words about the the gamma expansion models.
This helps us to fix the notations and introduce the basic equations needed here. Thus, let us consider the Hamiltonian
\begin{equation}
H=-{\frac{1}{\sin {3\gamma }}}{\frac{\partial }{\partial \gamma }}\sin {%
3\gamma }{\frac{\partial }{\partial \gamma }}+U+W,  \label{1}
\end{equation}
where $U$ is a periodic function in $\gamma$ with the period equal to $2\pi$ and\begin{equation}
W={\frac{1}{4}}\sum_{k=1}^{3}{\frac{1}{\sin ^{2}{(\gamma -{\frac{2\pi }{3}}k)%
}}}Q_{k}^{2}  \label{2}
\end{equation}
with $Q_k$ denoting the components of the intrinsic angular momentum.
 Any approximation for the potential term, for example by expanding it in
 power series of $\gamma$, alters the periodic behaviour of the eigenfunction
 unless a special caution is taken. Moreover, the approximating Hamiltonian
 loses its Hermiticity with respect to the scalar product defined with the
 measure for the gamma variable, $|\sin 3\gamma|d\gamma$.

We illustrate this by considering the case of a little more complex potential 
\begin{equation}
U=u_1\cos 3\gamma +u_2\cos ^{2} 3\gamma .  \label{9}
\end{equation}

Substituting $\varphi =
\sqrt{\left| \sin 3\gamma \right| }\psi $, the eigenvalue equation  $H\psi =E\psi $, becomes $\tilde{H}\varphi =0$ with

\begin{equation}
\tilde{H}=\frac{\partial ^{2}}{\partial \gamma ^{2}}+\frac{9}{4}\left[ 1+%
\frac{1}{\sin ^{2} 3\gamma }\right] -U-W+E.  \label{8}
\end{equation}

Suppose that $\left| \gamma \right| \ll 1$.    Expanding the terms in $\gamma$, in power series up to the fourth order, one obtains:
\begin{eqnarray}
U_{4}&=&u_1+u_2-9\gamma ^{2}\left( \frac{u_1}{2}+u_2\right) +27\gamma ^{4}\left( 
\frac{u_1}{8}+u_2\right) ,\nonumber  \\
W_{4} &=&\frac{1}{3}\left( 1+2\gamma ^{2}+\frac{26\gamma ^{4}}{9}\right)
\left( Q_{1}^{2}+Q_{2}^{2}\right)\\ 
&&+\frac{2\sqrt{3}\gamma }{9}
 \left(1+2\gamma ^{2}\right) \left( Q_{2}^{2}-Q_{1}^{2}\right)   \label{16}
\nonumber \\
&&+\frac{1}{4}\left( \frac{1}{\gamma ^{2}}+\frac{1}{3}+\frac{\gamma ^{2}}{15}%
+\frac{2\gamma ^{4}}{189}\right) Q_{3}^{2}.  \nonumber
\label{19}
\end{eqnarray}
The low index accompanying $U$ and $W$ suggests that the expansion was
truncated at the fourth order. Note that due to the term $W$, the equations of
motion for the variable $\gamma$ and Euler angles are coupled together. Such a coupling can in principle be handled as we did for the harmonic liquid drop in
Ref. \cite{Ghe1,Rad2}. 
Here,  we separate the equation for $\gamma$ by averaging $W_4$ with an
eigenfunction for the intrinsic angular momentum squared.
 The final result for $H_4$ is:
\begin{eqnarray}
H_{4} 
&=&\frac{\partial ^{2}}{\partial \gamma ^{2}}+\frac{1}{4\gamma ^{2}}\left(
1-\left\langle Q_{3}^{2}\right\rangle \right) +h_{0}+h_{2}\gamma
^{2}+h_{4}\gamma ^{4}\nonumber\\
&&+\frac{2\sqrt{3}\gamma }{9}\left( 1+2\gamma ^{2}\right)
\left\langle Q_{2}^{2}-Q_{1}^{2}\right\rangle ,  \\
h_0&=&E -\frac{1}{3}L(L+1)+\frac{1}{4}\left\langle Q^2_3\right\rangle 
-(u_0+u_1+u_2)+\frac{15}{2},
\nonumber\\
h_2&=&-\frac{2}{3}L(L+1)-\frac{13}{20}\left\langle Q^2_3\right\rangle+
\frac{9}{2}u_1+9u_2+
\frac{27}{20},\nonumber\\
h_4&=&-\frac{26}{27}L(L+1)-\frac{121}{126}\left\langle Q^2_3\right\rangle -\frac{27}{8}u_1-27u_2+\frac{27}{14}\nonumber.
\end{eqnarray}
where $L$ denotes the angular momentum.
If the average is made with the Wigner function $D^L_{MK}$, important simplifications are obtained since the following relations hold:
\begin{equation}
\left\langle Q_{2}^{2}-Q_{1}^{2}\right\rangle =0\;,\;\;
\left\langle Q_{3}^{2}\right\rangle =K^2\; .
\end{equation}
Let us stick to this situation.
Note that $H_4$ contains a singular term in $\gamma$ at, $\gamma=0$, coming
from the term coupling the intrinsic variable $\gamma$ with the Euler angles.
One could get rid of such a coupling term by starting with a potential in
gamma containing a singular term which cancels the contribution produced by
the $W$ term. Thus, the new potential would be

\begin{equation}
U'=U+\frac{9K^2}{4\sin^2 3\gamma}\;.
\label{uprime}
\end{equation}
The corresponding fourth order expansion of $\tilde{H}$ is:

\begin{eqnarray}
H^{\prime}_{4} 
&=&\frac{\partial ^{2}}{\partial \gamma ^{2}}+\frac{1}{4\gamma ^{2}} +h^{\prime}_{0}+h^{\prime}_{2}\gamma
^{2}+h^{\prime}_{4}\gamma ^{4} ,\\
h^{\prime}_0&=&h_0+K^2,\;
h^{\prime}_2=h_2+\frac{27}{20}K^2,\;
h^{\prime}_4 = h_4+\frac{27}{14}K^2\nonumber.
\end{eqnarray}
Some remarks concerning the equation $H^{\prime}_{4}\varphi=0$  are worth to be mentioned. If in this equation one ignores the $\gamma^4$ term, the resulting equation has the Laguerre functions as solutions and moreover the Hamiltonian
exhibits the $X(5)$ symmetry. Note also that the Hamiltonian coefficients are
different from those of Refs.[13,19]. The difference is caused by the fact that
here, the expansion $W_4$ is complete.
Taking in the expanded potential $u_1=u_2=0$ and ignoring, for $\gamma$ small,
the term $27K^2\gamma^2/20$, the resulting potential is that of an infinite
square well which was treated by Iachello in Ref. \cite{Iache9}. The solutions are, of course, the Bessel functions of half integer indices. None of the mentioned solutions is periodic. Also  the approximated Hamiltonians are not Hermitian in the Hilbert space of functions in gamma with the integration measure as introduced by LD. \

To overcome this principle problems,  we try first to avoid making
approximations. Thus, let us consider the Hamiltonian given by Eq. (\ref{1})
where instead of $U$ we consider  $ U^{\prime}$ as defined by Eq. \ref{uprime}
and moreover ignore $W$.  Changing the  variable $x=\cos3\gamma$,  the eigenvalue equation associated to this Hamiltonian becomes:
\begin{eqnarray}
&& \left( 1-x^{2}\right) \frac{{\rm d}^{2}S}{{\rm d}\,x^{2}}-
2x\frac{{\rm d} S}{{\rm d}\,x}+ \\
&&\left[\frac{1}
{9}(E-u_{1}x-u_{2}x^{2})
-\frac{K^{2}}{4{(1-}x^{2}{)}}\right] S=0.\nonumber
\label{9}
\end{eqnarray}
If $u_1=u_2=K=0$, the solution is the Legendre polynomial $P_n$ while $E=9n(n+1)$. Te case $u_1=u_2=0$ was considered recently in Ref.\cite{BFHH}.
For other particular choice of the coefficients $ u_1$ and $u_2$, the solution is readily obtained if one compares the above equation with that characterising the spheroidal oblate functions \cite{Abra}
\begin{eqnarray}
&&\left( 1-x^{2}\right) \frac{{\rm d}^{2}S_{nm}}{{\rm d}\,x^{2}}-
2x\frac{{\rm d}S_{nm}}
{{\rm d}\,x}\\
&+&(\lambda
_{nm}-c^{2}x^{2}-\frac{m^{2}}{{1-}x^{2}}) S_{nm}=0. \nonumber
 \label{10}
\end{eqnarray}
The prolate case is reached by changing $c\to {\rm i}c$.
 For $c=0$, the solutions of  Eq. (\ref{9}) are the associated Legendre functions
$P_{n}^{m}$. For $c\ne 0$, $S_{nm}$, with $m,n$ integers and $n \ge m\ge 0$,
are linear series of these functions.

In the case $u_{1}=0$, the solution of Eq. (10) is identified as being the spheroidal function while  the energy is  related to $\lambda_{mn}$ by a simple equation:
\begin{equation}
m=\frac{K}{2},\quad c^{2}=\frac{u_{2}}{9},\quad \lambda _{nm}=\frac{E}{9}\;.
\end{equation}
For $|c|$ small the energies $E_{nm}$ exhibit the asymptotic expansion
\begin{eqnarray}
E_{nm}&\approx &9n(n+1)-\frac{2\left[n(n+1)+m^2-1\right]}{(2n-1)(2n+3)}u_2
\label{12}\\
&+&\frac{1}{18}\frac{\left[(n-1)^2-m^2\right](n^2-m^2)}{(2n-3)(2n-1)^3(2n+1)}
u^2_2\nonumber\\
&-&\frac{1}{18}\frac{\left[(n+1)^2-m^2\right]\left[(n+2)^2-m^2\right]}
{(2n+1)(2n+3)^3(2n+5)}u^2_2.
\nonumber
\end{eqnarray}
Eq. (\ref{12}) considered for a fixed $m$ but various $n$, defines a band.
Similar expansions may be derived for $|c|$ large. In order to save the space
we give the $O(c^{-4})$ expansions for a few $E_{nm}$'s.
\begin{eqnarray}
E_{11}&\approx &9\left(\frac{1}{4}-c^2+c+\frac{5}{16c}+\frac{33}{64c^2}\right),\nonumber\\
E_{21}&\approx &9\left(-\frac{3}{4}-c^2+3c+\frac{9}{16c}+\frac{135}{64c^2}\right),\nonumber\\
E_{22}&\approx &9\left(\frac{13}{4}-c^2+c+\frac{29}{16c}+\frac{177}{64c^2}\right),\nonumber\\
E_{31}&\approx &9\left(-\frac{11}{4}-c^2+5c-\frac{5}{16c}+\frac{219}{64c^2}\right),\nonumber\\
E_{32}&\approx &9\left(\frac{9}{4}-c^2+3c+\frac{81}{16c}+\frac{855}{64c^2}\right),\nonumber\\
E_{33}&\approx &9\left(\frac{33}{4}-c^2+c+\frac{69}{16c}+\frac{417}{64c^2}\right).
\end{eqnarray}
Now, we shall focus on an approximate solution which preserves the periodicity in $\gamma$. To this aim, we consider the Hamiltonian
\begin{eqnarray}
H&=&-\frac{1}{\sin 3\gamma}\frac{\partial}{\partial \gamma}\sin 3\gamma \frac{\partial}{\partial \gamma}+U(\gamma),\nonumber\\
U(\gamma) &=&u_1\cos3\gamma+u_2\cos^23\gamma+\frac{A}{4\sin^2\gamma}.
\end{eqnarray}
with A a positive constant.
Changing the function by the transformation 
$\Psi=|\sin 3\gamma |^{-1/2}{\Phi}$, for $\sin(3\gamma )\ne 0$, the eigenvalue equation for H becomes:
\begin{equation}
\left[\frac{\partial^2}{\partial \gamma^2}+E+\frac{9}{4}+\frac{9}{4\sin^23\gamma}-U(\gamma)\right]{\Phi}=0.
\label{25}
\end{equation}
Under the regime of $|\gamma|$ small, we take the $O(\gamma ^3)$ expansion of
the terms depending on $\gamma$ and in the final expression
approximate $\gamma \approx \sin\gamma$. In this way, Eq. \ref{25} becomes:

\begin{eqnarray}
&&\left(\frac{\partial^2}{\partial \gamma^2}+a-2q\cos2\gamma-\frac{K^2-1}{4\sin^2\gamma}\right){\Phi}=0,\;\rm{with}\label{26}\\
&&q=\frac{1}{3}+\frac{9}{8}u_1+\frac{9}{4}u_2,u=u_2+\frac{347}{108},a=E+\frac{10}{9}q+u\nonumber .
\end{eqnarray}
We suppose now that this equation is valid in the interval $[0,2\pi]$.
The equation (\ref{26}) is just the trigonometric form of the spheroidal functions. The algebraic version is obtained  by changing the variable $x=\cos\gamma.$
For $A=1$, one obtains the Mathieu equation:
\begin{equation}
\left(\frac{\partial^2}{\partial \gamma^2}+a-2q\cos2\gamma\right)\Phi=0.
\end{equation}
There are two sets of solutions, one even and one odd denoted by $\Phi^+(a,q,\gamma)$ and $\Phi^-(a,q,\gamma)$, respectively. For $q=0$, both solutions are periodic, irrespective of $a$.
\begin{equation}
\Phi^+(a,0,\gamma)=\cos\left(\sqrt{a}\gamma\right),\;\Phi^-(a,0,\gamma)=
\sin\left(\sqrt{a}\gamma\right).
\end{equation}

\begin{figure}[h!]
\begin{center}
\includegraphics[height=6cm]{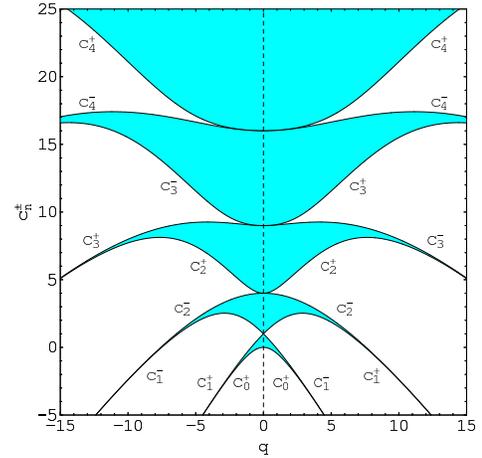}
\end{center}
\caption{The characteristic curves $c^{\pm}_n$  are plotted as functions of $q$
 for several values of $n$.}
\label{Fig. 1}
\end{figure}
For $q\ne 0$, the Mathieu functions are periodic in $\gamma$ only for a
certain set of values of $a$, called characteristic values. These are denoted
by $c^+_n$ for even and $c^-_n$ for odd functions, respectively. In the
plane ($a,q$), the characteristics curves $c^{\pm}_n$   separate the stability
regions, shown in Fig.1 by black colour, from the nonstability  ones, indicated by white colour in the quoted figure.
For $q=0$ the equalities $c^{\pm}_n(0)=n^2$ hold.
By means of Eq. (\ref{26}) the characteristic values determine the energy $E$.
Thus, the energy spectrum is given by $E^{\pm}_n-u$ with $E^{\pm}_n=c^{\pm}_n-\frac{10}{9}q$. The corresponding wave functions are the elliptic cosine and elliptic sine functions respectively:
\begin{eqnarray}
 \Phi^+_0(q,\gamma)&=&\frac{1}{\sqrt{2\pi}}ce_0(q,\gamma),
\Phi^+_n(q,\gamma)=\frac{1}{\sqrt{\pi}}ce_n(q,\gamma),\nonumber\\
\Phi^-_n(q,\gamma)&=&\frac{1}{\sqrt{\pi}}se_n(q,\gamma),\;n=1,2,...
\end{eqnarray}
They form an orthogonal set. The matrix elements of the gamma depending factors of the transition operator can be easily calculated in Mathematica. Moreover, in the regime of $|q|$-small, these matrix elements can be analytically calculated, since the following representation of the wave functions hold:
\begin{eqnarray}
&&\Phi^{\pm}_n(\gamma)\approx \cos(n\gamma-\theta_{\pm})-\\
& &\left[\frac{\cos\left[(n+2)\gamma-\theta_{\pm}\right]}{4(n+1)}-
\frac{\cos\left[(n-2)\gamma-\theta_{\pm}\right]}{4(n-1)}\right]q^2.\nonumber
\end{eqnarray}
where $n\ge 3, \theta_{+}=0$ and $\theta_-=\pi/2$. The corresponding energies have the following expressions:
\begin{eqnarray}
E^+_0&\approx& u-\frac{10}{9}q-\frac{q^2}{2},
E^+_1\approx E^-_1\approx u -\frac{10}{9}q-\frac{q^2}{8},\nonumber\\
E^+_2&\approx& E^-_2\approx u+4-\frac{10}{9}q-\frac{q^2}{2},\\
E^+_n&\approx& E^-_n\approx u+n^2-\frac{10}{9}q+\frac{q^2}{2(n^2-1)},\;n\ge 3.
\nonumber
\end{eqnarray}

Similar expansions can be derived also for $|q|\gg 1$. Here we give only the energy expressions:
\begin{eqnarray}
E^+_n & \approx & E^-_{n+1}\\
& \approx &-\frac{28}{9}q+2(2n+1)\sqrt{q}-\frac{1}{4}(2n^2+2n+1).\nonumber
\end{eqnarray} 

Obviously, a phase transition is determined by the combined effects coming from
the behaviour of the wave function in the $\beta$ and $\gamma$ variables, respectively.
For the $X(5)$ symmetry, the $\beta$ variable is described by a Bessel function of irrational index, while $\gamma$ by a Laguerre polynomial. Here, we propose to change the description of the $\gamma$ variable either by a spheroidal or by a Mathieu function. These functions are periodic and the corresponding Hamiltonians Hermitian. Moreover, in both versions, the $X(5)$ Hamiltonian is obtained in the limit of small $|\gamma |$.

\end{document}